\documentstyle[fleqn,12pt,aps]{revtex}
\input amssym.def
\begin{document}
\baselineskip=18pt
\begin{center}
{\bf Gauge Theory Without Ghosts}
\bigskip

\bigskip

\bigskip

Bryce DeWitt and Carmen Molina-Par\'{\i}s\\
Center for Relativity, Department of Physics, University of Texas,\\
Austin, Texas  78712--1081
\end{center}

\bigskip

\bigskip

\noindent {\bf Abstract}
\medskip

A quantum effective action for gauge field theories is constructed that
is gauge invariant and independent of the choice of gauge breaking terms
in the functional integral that defines it. The loop expansion of this
effective action leads to new Feynman rules, involving new vertices but
without diagrams containing ghost lines. The new rules are given
in full for pure Yang-Mills theory, and renormalization
procedures are sketched. No BRST arguments are needed. With the new
rules the $\beta$ function becomes ghost independent to all orders.
Implications for a stably-based ghost-free attack on the back-reaction
problem in quantum gravity are briefly discussed.


\pagebreak

The purpose of this Letter is to show that by using the geometry of the
space-of-histories $\Phi$ of a gauge field in a judicious way one can
develop a renormalizable perturbation theory in which ghosts play no
role.

$\Phi$ is a principal fibre bundle having for its typical fibre the
gauge group $\cal G$. Real physics takes place in the base space
$\Phi/\cal G$. Since $\cal G$ is a group manifold it admits a group
invariant Riemannian metric. This metric can be extended in an infinity
of ways to an invariant metric $\boldmath \gamma$ on $\Phi$, but it
turns out that if the extended metric is required to be {\it ultralocal}
then, up to a scale factor, it is unique in the case of Yang-Mills fields
and belongs to a one-parameter family in the case of gravity.

The classical action $S$ is invariant under gauge transformations, which
have the infinitesimal form
\begin{eqnarray}
\delta \varphi^{i} = {Q^{i}}_{\alpha} \delta \xi^{\alpha},
\end{eqnarray}
the $\varphi^{i}$ being the fields (which are to be viewed as
coordinates in $\Phi$) and the $\delta \xi^{\alpha}$ being infinitesimal
gauge parameters. Here all indices play a double role, specifying
discrete labels as well as points of spacetime, which means that
summations over repeated indices include integrations over spacetime.

Group invariance of $\boldmath \gamma$ is the statement
\begin{eqnarray}
{\pounds_{\bf Q_{\alpha}}} \mbox{\boldmath $\gamma$} = 0,
\end{eqnarray}
where the $\bf Q_{\alpha}$ are vector fields on $\Phi$ having the
components $Q^{i}_{\alpha}$. The ${\bf Q}_{\alpha}$ are
{\it Killing vector fields} for $\mbox{\boldmath $\gamma$}$
and {\it vertical vector fields} for the principal bundle $\Phi$.
The $\bf Q_{\alpha}$ and $\boldmath \gamma$ together
define a unique
connection 1-form on $\Phi$:
\begin{eqnarray}
\mbox{{\boldmath $\omega$}}^{\alpha} = {\frak G}^{\alpha \beta} \bf {Q}_{\beta}
\cdot
\mbox{\boldmath $\gamma$}
\end{eqnarray}
where ${\frak G}^{\alpha \beta}$ is the Green's function, appropriate to the
boundary
conditions at hand in a given problem, of the operator
\begin{eqnarray}
{\frak F}_{\alpha \beta} = -\bf {Q}_{\alpha} \cdot \mbox{\boldmath $\gamma$}
\cdot \bf{Q}_{\beta}.
\end{eqnarray}
It is easy to see that $ {\mbox{\boldmath $\omega$}}^{\alpha} \cdot
\bf{Q}_{\beta} =
 {\delta^{\alpha}}_{\beta}$ and that horizontal vectors are those that
are perpendicular to the fibres (under the metric
$\mbox{\boldmath$\gamma$}$). A horizontal vector may be obtained
from any vector by application of the {\it horizontal projection operator}:
\begin{eqnarray}
{\Pi^{i}}_{j} = {\delta^{i}}_{j} - {Q^{i}}_{\alpha} {\omega^{\alpha}}_{j}.
\end{eqnarray}

The trick for avoiding ghosts is to make use of the following connection
on the frame bundle $F \Phi$:
\begin{eqnarray}
{\Gamma^{i}}_{jk} & = & {\Gamma_{\gamma}}^{i}{_{jk}}
- {Q^{i}}_{\alpha \cdot j} {\omega^{\alpha}}_{k}
- {Q^{i}}_{\alpha \cdot k} {\omega^{\alpha}}_{j} \nonumber \\
& & + \frac{1}{2} {\omega^{\alpha}}_{j} {Q^{i}}_{\alpha \cdot l}
{Q^{l}}_{\beta} {\omega^{\beta}}_{k} + \frac{1}{2}
{\omega^{\alpha}}_{k} {Q^{i}}_{\alpha \cdot l}
{Q^{l}}_{\beta} {\omega^{\beta}}_{j}.
\end{eqnarray}
Here $\Gamma_{\gamma}$ is the Riemannian connection associated
with $\boldmath \gamma$ and the dots denote covariant differentiation
based on it. The connection (6), which was first introduced by G. A.
Vilkovisky $[1]$, has the following remarkable properties:
%\begin{itemize}
%\item
\medskip

\noindent 1. {Let $\lambda$ be a geodesic based on it. If the tangent
vector to $\lambda$ is horizontal at one point along
$\lambda$ then (a) it is horizontal everywhere along $\lambda$, (b)
$\lambda$ is also a geodesic based on $\Gamma_{\gamma}$, and
(c) $\lambda$ is the horizontal lift of a Riemannian geodesic in
$\Phi/{\cal G} $ based on the natural projection of $\boldmath \gamma$
down to $\Phi / \cal G$ (which exists because of the group invariance of
$\boldmath \gamma).$}
%\item
\medskip

\noindent {2. Alternatively, if $\lambda$ is tangent to a fibre at one point
then it lies in that fibre.}
%\item
\medskip

\noindent {3. For all $\alpha$
\begin{eqnarray}
{Q^{i}}_{\alpha;j} = \frac{1}{2}{Q^{i}}_{\gamma}{c^{\gamma}}
_{\alpha \beta} {\omega^{\beta}}_{j},
 \end{eqnarray}
where the semicolon denotes covariant functional differentiation based
on Vilkovisky's connection and the ${c^{\gamma}}_{\alpha \beta}$ are the
structure constants of $\cal G$.}
%\item

\noindent4. {If $A$ is a gauge invariant functional (e.g., $A = S$) then for
all $n$
\begin{eqnarray}
A_{;(i_{1} \ldots i_{n})} = A._{(j_{1} \ldots j_{n})}
\Pi^{j_{1}}{_{i_{1}}} \ldots \Pi^{j_{n}}{_{i_{n}}},
\end{eqnarray}
where the parentheses indicate that a symmetrization of the indices they
embrace is to be performed.}
%\item
\medskip

\noindent {5. If one introduces a fibre-adapted coordinate system $(I^{A},
K^{\alpha})$, where the $I^{A}$ label the fibres (and are gauge
invariant) and the $K^{\alpha}$ label the points in each fibre, then the
components of the $nth$ covariant functional derivative of any gauge
invariant quantity vanishes unless all indices are capital Latin.}
%\end{itemize}
\bigskip

In practice the $I^{A}$ are purely conceptual, but the $K^{\alpha}$ need to
be explicitly chosen. One must single out a {\it base point}
$\varphi_{*}$ in $\Phi$ and choose the $K$'$s$ so that the matrix
\begin{eqnarray}
{\hat{\frak F}}^{\alpha}{_{\beta}} = {\bf Q_{\beta}} K^{\alpha}
\end{eqnarray}
is a nonsingular differential (or pseudodifferential) operator at and in
a neighborhood of $\varphi_{*}$. Typical convenient choices for
$\varphi_{*}$ are ${A^{\alpha}}_{\mu}(x)_{*} = 0$ in Yang-Mills theory and
$g_{\mu \nu}(x)_{*} = \eta_{\mu \nu}$ (Minkowski metric) in gravity
theory. It turns out to be extremely useful to use the base point also
for another purpose.

Let $\varphi$ be an arbitrary point of $\Phi$ and $\lambda$ a geodesic
connecting it to the base point. Let $s$ and $s_{*}$ be the values, at
$\varphi$ and $\varphi_{*}$ respectively, of an affine parameter along
$\lambda$. Define
\begin{eqnarray}
\mbox{\boldmath $\phi$} = (s -s_{*}) {\left (\frac{\partial}
{\partial s} \right )}_{\lambda (s_{*})}.
\end{eqnarray}
$\mbox{\boldmath $\phi$}$ is a vector at $\varphi_{*}$, invariant
under rescaling of s. Its components $\phi^{a}$ in an arbitrary frame
at $\varphi_{*}$ may be called {\it Gaussian normal fields}.
Vilkovisky $[1]$ has shown how the $\phi^{a}$
can be used to carry out covariant
functional Taylor expansions about $\varphi_{*}$. Such expansions are
important in the theory of the quantum effective action by virtue of the
fact that, in a fibre-adapted coordinate frame, $\phi^{A}$ depends only
on the $I^{A}_{*}$ and $I^{A}$.

One begins by writing
\begin{eqnarray}
e^{i \Gamma [I_{*}, \bar{I}]} = N \int e^{i S [I_{*}, I]
+ i \Gamma_{1}[I_{*}, \bar{I}] (\bar{I} - I)} \mu [I_{*}, I] [d I].
\end{eqnarray}
where $\mu$ is an appropriate measure functional and the quantum
effective action $\Gamma$ is understood to be obtainable (in principle)
by an iterative procedure based on computing a loop expansion for the
functional integral. Here the $I^{A}$ are assumed to be already
Gaussian normal (i.e., they are the $\phi^{A}$) so that the
difference $\bar{I} - I$ is a vector at $I_{*}$ and makes
good geometrical sense when contracted with the functional derivative
(gradient)
$\Gamma_{1}$ of $\Gamma$ with respect to $\bar{I}$.

Although eq. (11), which recognizes that real physics takes place in
$\Phi / \cal G$, is a reasonable starting
point, it is of formal validity only because the $I^{A}$ are purely
conceptual. For explicit calculations one must pass from
$\Phi/{\cal G} $ to $\Phi$ by introducing
variables $K^{\alpha}$. These too may be chosen Gaussian normal, in
which case both the $I^{A}$
and the $K^{\alpha}$ are necessarily linearly related to the $\phi^{a}$
in an arbitrary frame at $\varphi_{*}$:
\begin{eqnarray}
I^{A} = {P^{A}}_{a}[\varphi_{_{*}}]\phi^{a}, \; \; \; \; \; \;
K^{\alpha} = {P^{\alpha}}_{a}[\varphi_{*}]\phi^{a}.
\end{eqnarray}
($\varphi_{*}$ can now be any point in the fibre over $I_{*}$.) Equation
(11) can then be rewritten with a Gaussian gauge breaking term thrown
in:
\begin{eqnarray}
e ^{i \Gamma [I_{*}, \bar{I}]}
& = & N \int
[d I] \int [d K]
e^{i (S [I_{*}, I] + \frac{1}{2} \kappa_{\alpha \beta}[\varphi_{*}] K^{\alpha}
K^{\beta}) + i \Gamma_{1} [I_{*}, \bar{I}] (\bar{I} -I)} \nonumber \\
& & \; \; \; \; \; \; \; \; \; \; \; \; \; \; \; \; \; \; \; \; \; \;
\times  ({\rm det} \kappa [\varphi_{*}])^{\frac{1}
{2}} \mu [I_{*}, I].
\end{eqnarray}
Since $\phi^{A} \; (= I^{A})$ depends only on the $I^{A}_{*}$ and $I^{A}$
one can immediately transform to Gaussian normal fields in an arbitrary
frame, noting, by virtue of (12), that the Jacobian $J[\varphi_{*}]$ for
the transformation is a {\it constant}. We shall abuse notation slightly
by writing $\Gamma[I_{*},\bar{I}] = \Gamma[\varphi_{*}, \bar{\phi}]$ where
the ${\bar \phi}^{a}$ are such that $\bar{I}^{A} =
{P^{A}}_{a}[\varphi_{_{*}}]\bar{\phi}^{a}$. We stress that, despite the
presence of the gauge breaking term, $\Gamma[\varphi_{*}, \bar{\phi}]$ is
still gauge invariant and independent of the choice of $P$'$s$ and
$\kappa$'$s$, provided the $\bar{\phi}^{a}$ are held at values such that
$\bar{K}^{\alpha} = {P^{\alpha}}_{a}[\varphi_{*}]\bar{\phi}^{a}$ vanishes.
However, if the $\bar{\phi}^{a}$ are allowed to run free then $\Gamma$
suffers merely the simple modification $\Gamma \rightarrow \hat{\Gamma}$ where
\begin{eqnarray}
\hat{\Gamma}[\varphi_{*}, \bar{\phi}] = \Gamma [\varphi_{*}, \bar{\phi}]
+  \frac{1}{2} \kappa_{\alpha \beta}[\varphi_{*}] {P^{\alpha}}_{a}
[\varphi_{*}]{P^{\beta}}_{b}[\varphi_{*}] \bar{\phi}^{a} \bar{\phi}^{b}.
\end{eqnarray}
Hence finally
\begin{eqnarray}
e^{i \hat{\Gamma}[\varphi_{*}, \bar{\phi}]}
& = & N \int
e^{i (S [\varphi_{*},{\phi}] + \frac{1}{2} \kappa_
{\alpha \beta} [\varphi_{*}] {{P}^{\alpha}}_{a}[\varphi_{*}]
{P^{\beta}}_{b}[\varphi_{*}]\phi^{a} \phi^{b})
+ i \hat{\Gamma}_{1}[\varphi_{*}, \bar{\phi}] (\bar{\phi} - \phi)} \nonumber \\
& & \; \; \; \; \; \; \times ( \rm{det} \kappa [\varphi_{*}])^{\frac{1}{2}}
J[\varphi _{*}]
\mu [\varphi_{*}, \phi] [d \phi].
\end{eqnarray}

Now let ${\hat{\frak G}^{\alpha}}{_{\beta}}$ be the Green's function of the
operator ${\hat{\frak F}^{\alpha}}{_{\beta}}$ of eq. (9). It is easy to show
that $J{\rm det} \hat{\frak G}$ is independent of the choice of the
$K$'$s$ (i.e., of the ${{P}^{\alpha}}_{a}$). Hence, as far as
assuring the $P$-independence of $\Gamma$ is concerned, $J$ may be
replaced by ${({\rm det}{\hat{\frak G}})}^{-1}$ in the functional
integral. This is the well known ghost determinant. It is not difficult
to obtain a covariant functional Taylor expansion of
$ln{\rm det}\hat{\frak G}$, and to use it to compute vertices for the
interaction of ghosts with the fields $\phi^{a}$. These vertices turn
out to have such a form as to conspire to cause every Feynman graph
containing a ghost line to vanish. It is clear, from the constancy of
$J$, that the ghost can play no real diagrammatic role. It has become,
as it were, a ghost of itself.

In what follows we shall drop $\rm{det} \kappa$, $J$ and the measure from the
integrand of eq. (15). (It is shown in reference $[2]$ that the chief role
of the measure is to justify throwing away nonvanishing contributions
from arcs at infinity in the Wick rotation procedure.) The loop
perturbation series is then obtained by expanding the integrand about
$\bar{\phi}$, writing
\begin{eqnarray}
\phi = \bar{\phi} + \chi
\end{eqnarray}
and using
\begin{eqnarray}
S[\varphi_{*}, \bar{\phi} + \chi] & = &  \sum_{n = 0}^{\infty}
\frac{1}{n!} S,_{a_{1} \ldots a_{n}}[\varphi_{*}, \bar{\phi}]
\chi^{a_{1}} \ldots \chi^{a_{n}},
\end{eqnarray}
where the ordinary derivatives of $S [\varphi_{*}, \bar{\phi}]$ with
respect
to the $\bar{\phi}^{a}$ are given by
\begin{eqnarray}
S,_{a_{1} \ldots a_{n}} [\varphi_{*}, \bar{\phi}]
& = & \sum_{m = 0}^{\infty} \frac{1}{m!}
S_{;(a_{1} \ldots a_{n} b_{1} \ldots b_{m})} [\varphi_{*}]
\bar{\phi}^{b_{1}} \ldots \bar{\phi}^{b_{m}} .
\end{eqnarray}
The loop graphs themselves are embodied in the functional
\begin{eqnarray}
\Sigma [\varphi_{*}, \bar{\phi}]
& = & \Gamma [\varphi_{*}, \bar{\phi}] - S[\varphi_{*}, \bar{\phi}]
\end{eqnarray}
which is generated by the functional integral equation (alternative to (15))
\begin{eqnarray}
e^{i \Sigma[\varphi_{*}, \bar{\phi}]}
= N \int e^{i(-\Sigma_{1} [\varphi_{*}, \bar{\phi}] \chi + \frac{1}{2} F
[\varphi_{*}, \bar{\phi}] \chi \chi + \frac{1}{6} S_{3}[\varphi_{*},
\bar{\phi}] \chi \chi \chi + \ldots)}[d \chi]
\end{eqnarray}
where
\begin{eqnarray}
F[\varphi_{*},\bar{\phi}] & = & S_{2} [\varphi_{*},\bar{\phi}] +
{{P}^{tr}}[\varphi_{*}] \kappa[\varphi_{*}] P[\varphi_{*}].
\end{eqnarray}
Amplitudes for physical processes that take place in the background
$\varphi_{*}$ are obtained by attaching external lines to the graphs and
setting $\bar{\phi} = 0$. The vertex functions (18) then reduce (see eq. (8))
to
 \begin{eqnarray}
S,_{a_{1} \ldots a_{n}} [\varphi_{*}, 0] = S._{{(b_{1} \ldots b_{n})}}
[\varphi_{*}] \Pi^{b_{1}}{_{a_{1}}} [\varphi_{*}] \ldots \Pi^{b_{n}}{_{a_{n}}}
[\varphi_{*}],
\end{eqnarray}
and these are easily calculated. The secret of ghost-free
gauge theory is seen to be very simple:
Replace all traditional vertex functions by (22) and throw
away the ghost diagrams.

The new rules are particularly simple in the case of Yang-Mills theory.
The classical action, in a Minkowski spacetime of $N$ dimensions, is
\begin{eqnarray}
S = -\frac{1}{4} \int \gamma_{\alpha \beta}{F^{\alpha}}_{\mu \nu}
F^{\beta \mu \nu} d^{N}x,
\end{eqnarray}
where $\gamma_{\alpha \beta}$ is the Cartan-Killing metric of the associated
Lie algebra (assumed simple and non-Abelian) and
\begin{eqnarray}
{F^{\alpha}}_{\mu \nu} = {\partial}_{\mu}{A^{\alpha}}_{\nu} -
{\partial}_{\nu}{A^{\alpha}}_{\mu} + g_{0}{f^{\alpha}}_{\beta \gamma}
{A^{\beta}}_{\mu} {A^{\gamma}}_{\nu},
\end{eqnarray}
$g_{0}$ being the bare coupling constant and
${f^{\alpha}}_{\beta \gamma}$ the structure constants of the algebra,
related to $\gamma_{\alpha \beta}$ by
\begin{eqnarray}
{f^{\gamma}}_{\alpha \delta}{f^{\delta}}_{\beta \gamma} = -\lambda
{\gamma}_{\alpha \beta}
\end{eqnarray}
for some positive constant $\lambda$ that depends on the scale choice of
the algebra basis. The gauge transformation law (1) takes the form
\begin{eqnarray}
\delta {A^{\alpha}}_{\mu} = \int {Q^{\alpha}}_{\mu \beta'}\delta
\xi^{\beta'}d^{N}x'
\end{eqnarray}
where
\begin{eqnarray}
{Q^{\alpha}}_{\mu \beta'} = (- {\delta^{\alpha}}_{\beta}
{\partial}_{\mu}
+ g_{0} {f^{\alpha}}_{\beta \gamma} {A^{\gamma}}_{\mu}) \delta (x, x'),
\end{eqnarray}
and the structure constants of $\cal G$ are
\begin{eqnarray}
{c^{\alpha}}_{\beta' \gamma''} = g_{0}{f^{\alpha}}_{\beta \gamma}
\delta (x, x')\delta (x, x'').
\end{eqnarray}

The unique ultralocal invariant metric is
\begin{eqnarray}
{\gamma_{\alpha}}{^{\mu}}{_{\beta'}}{^{\nu'}} = \gamma_{\alpha \beta} \;
\eta^{\mu \nu} \delta (x, x'),
\end{eqnarray}
which, being constant, is flat. The Riemannian connection components
vanish in the coordinates ${A^{\alpha}}_{\mu}(x)$, and the dots in eqs.
(6), (8) and (22) denote ordinary functional derivatives. This means
that there are only two distinct vertex functions, $S_{3}$ and $S_{4}$,
just as in the traditional formalism.

The ${{P}^{\alpha}}_{a}$ and $\kappa_{\alpha \beta}$ for Yang-Mills
fields are conveniently chosen to be
\begin{eqnarray}
{P^{\alpha}}{_{\beta'}}^{\mu'} = - {\delta^{\alpha}}_{\beta} \;
\eta^{\mu \nu} {\partial _{\nu}} \delta (x, x'),
\end{eqnarray}
\begin{eqnarray}
\kappa_{\alpha \beta'} = - \gamma_{\alpha \beta} \; \delta (x, x').
\end{eqnarray}
When ${A^{\alpha}}_{\mu *} = 0$ and $\bar {\phi} = 0$ the operator (21)
then takes the simple form
\begin{eqnarray}
{F_{\alpha}}{^{\mu}}{_{\beta'}}^{\nu'} \rightarrow -\gamma_{\alpha \beta} \;
\eta^{\mu \nu} p^{2},
\end{eqnarray}
where ``$\rightarrow$'' means ``pass to the Fourier transform.''
Moreover, the horizontal projection operator (5) becomes
\begin{eqnarray}
{\Pi^{\alpha}}{_{\mu \beta'}}^{\nu'} \rightarrow
{\delta^{\alpha}}_{\beta}
({\delta_{\mu}}^{\nu} - p_{\mu} p^{\nu}/p^{2}).
\end{eqnarray}
It is easy to see that one may restate the calculational rules for
Yang-Mills theory as follows:
%\begin{enumerate}
%\item
\medskip

\noindent {1. Retain only those traditional graphs that contain no ghost lines
and, in
these, use the traditional vertices.}
%\item
\medskip

\noindent {2. For the internal lines use the Green's function in the Landau
gauge.}
%\item
\medskip

\noindent {3. Apply the projection operator (33) to all external prongs.}
%\end{enumerate}
\medskip

Renormalization in the ghost-free formalism, although technically requiring
as much work as in the traditional formalism, is conceptually simpler.
There are only two independent renormalization constants instead of three, a
wave
function renormalization constant $Z$ and a constant $Y$ that renormalizes
the three-pronged vertex. The constant, call it $X$, that renormalizes
the four-pronged vertex is fixed by gauge invariance to be $X = Z^{-1}Y^{2}$.
The renormalized fields ${A_{R}}^{\alpha}{_{\mu}}$ and renormalized coupling
constant $g$ are defined by
\begin{eqnarray}
{A^{\alpha}}_{\mu} = Z^{{1}/{2}} {A_{R}}^{\alpha}{_{\mu}},
\; \; \; \; \; \; \; \;  g_{0} = \mu^{2-N/2} Z^{-3/2} Y g,
\end{eqnarray}
where $\mu$ is the usual auxiliary mass. Because the Gaussian normal
fields are nonlocally related to the ${A_{R}}^{\alpha}{_{\mu}}$, it
turns out that there is an additional {\it nonlocal} gauge invariant
term, of the form
\begin{eqnarray}
{\Delta}S & = & {\Xi \over 24}
{tr}(f_{\alpha}f_{\beta}f_{\gamma}f_{\delta})
({\eta^{\mu \nu}}{\eta^{\sigma \tau}} + {\eta^{\mu \sigma}}
{\eta^{\nu\tau}} + {\eta^{\mu \tau}}{\eta^{\nu \sigma}}) \nonumber\\
& & \times \int {d^{N}x} \int {d^{N}x'}\int {d^{N}x''}
\int {d^{N}x'''}\int {d^{N}x''''} {{\Pi^{\alpha}}_{\mu
{\bar{\alpha}}'}}^{{\bar{\mu}}'}
{{\Pi^{\beta}}_{\nu {\bar{\beta}}''}}^{{\bar{\nu}}''}\nonumber\\
& & \times \;{{\Pi^{\gamma}}_{\sigma {\bar{\gamma}}'''}}^{{\bar{\sigma}}'''}
{{\Pi^{\delta}}_{\tau {\bar{\delta}}''''}}^{{\bar{\tau}}''''}
{{A_{R}}^{\bar{\alpha}}{}^{\prime}}_{{\bar {\mu}}^{\prime}}
{{A_{R}}^{\bar{\beta}}{}^{\prime \prime}}_{{\bar {\nu}}{}^{\prime \prime}}
{{A_{R}}^{\bar{\gamma}}{}^{\prime \prime \prime}}_{{\bar {\sigma}}{}^{\prime
\prime \prime}}
{{A_{R}}^{\bar{\delta}}{}^{\prime \prime \prime \prime}}_{{\bar
{\tau}}{}^{\prime \prime \prime \prime}},
\end{eqnarray}
$(f_{\alpha} =({f^{\beta}}_{\alpha \gamma}))$ that must be added to the
classical action, whose only role is to mop up some residues coming from
the divergent parts of diagrams having four external prongs. This term
makes no contribution to the particle content of the theory, plays no
role in determining the $\beta$ function, and, under dimensional
regularization with minimal subtraction, does not even have to be
computed.

Diagrams in the ghost-free formalism have the same degree of divergence
as in the traditional formalism, and counter terms are computed in the
usual way. For pure Yang-Mills theory one finds
\begin{eqnarray}
Z = 1 - {{25 \over 6} {\lambda \over {(4 \pi)}^{2}} {1 \over N-4}
\; g^{2}} + O({g}^{4}),
\end{eqnarray}
\begin{eqnarray}
Y = 1 - {{11 \over 4} {\lambda \over {(4 \pi)}^{2}} {1 \over N-4}
\; g^{2}} + O({g}^{4}),
\end{eqnarray}
\begin{eqnarray}
X = 1 - {{4 \over 3} {\lambda \over {(4 \pi)}^{2}} {1 \over N-4}
\; g^{2}} + O({g}^{4}),
\end{eqnarray}
\begin{eqnarray}
{\Xi} = {1 \over 6} {\lambda \over {(4 \pi)}^{2}} {1 \over N-4}
\; g^{2} + O({g}^{4}),
\end{eqnarray}
\begin{eqnarray}
\beta = {\mu} {dg \over d{\mu}} = - {7 \over 2}{\lambda \over {(4 \pi)}^{2}}
\; g^{3} + O(g^{5}).
\end{eqnarray}
Because the Gaussian normal fields are nonlinearly related to the
${A_{R}}^{\alpha}{_{\mu}}$ the relation between the renormalized
coupling constant $g$ and that of the traditional theory, call it $\hat
{g}$, is also nonlinear:
\begin{eqnarray}
g = {\sqrt{22/21}} \; {\hat g} + O({\hat g}^{3}).
\end{eqnarray}
This yields the more familiar $\beta$ function
\begin{eqnarray}
{\hat {\beta}} = \mu {{d \hat{g}} \over {d \mu}} = - {11\over 3}
{\lambda \over {(4 \pi)}^{2}}\; {{\hat g}^{3}} + O({\hat g}^{5}).
\end{eqnarray}

The Gaussian normal fields to which one is referring here are simply the
Gaussian normal fields based on ${A^{\alpha}}_{\mu *} = 0$ and viewed in
the coordinate system defined by the ${A_{R}}^{\alpha}{_{\mu}}$. It is
natural to denote them by ${{\phi}_{R}}^{\alpha}{_{\mu}}$. The obstacle
that for years has  prevented theorists from looking at Yang-Mills
theory in terms of the ${{\phi}_{R}}^{\alpha}{_{\mu}}$ is the fact that
the ${{\phi}_{R}}^{\alpha}{_{\mu}}$ are nonlocally related to the
${{A}_{R}}^{\alpha}{_{\mu}}$. But the nonlocality in the neighborhood of
${\bar \phi}_{R} = 0$ is effectively just that produced by the
projection operator (33) and in the end presents no problem at all.

The ultimate importance of the ghost-free formalism lies not in its
demonstration that one can actually get along without ghosts (and the
whole paraphernalia of BRST as well!), thus remaining close to the
spirit of the classical theory from which one starts. What is important
is the fact that one obtains a quantum effective action $\Gamma$ that is
independent of the choice of gauge breaking terms and of the ghost
propagator (although not independent of the choice of the base point
$\varphi_{*}$).
This has implications for attempts to tackle not only problems with
``in-out'' boundary conditions, which constitute the usual framework for
the quantum effective action, but also problems involving ``in-in''
boundary conditions, in which one tries to get information about
expectation values. By using some variant of the well-known ``two-time
formalism'', together with Vilkovisky's ideas, one should be able to
construct an ``in-in'' effective action that is ghost and gauge-breaking
independent. This would lead to the possibility of finding
gauge-covariant, ghost-independent nonlocal dynamical equations that
govern, at least approximately, the evolution of field expectation
values.

The most interesting application of such a development would be in
quantum gravity. Here, of course, there are formidable obstacles: (a)
The algebraic complexity that always arises in gravitational
calculations. (b) The fact that the number of basic vertices is no
longer finite. (c) The fact that $\boldmath \gamma$ is no longer flat.
(d) The fact that gravity is not perturbatively renormalizable.
Nevertheless it would be both useful and feasible to examine the
``in-out'' and ``in-in'' one-loop quantum gravity effective actions, in
ordinary space and in arbitrary backgrounds. This is a minimum
requirement, for example, for setting up a stably-based
ghost-independent attack on the black-hole back-reaction problem.

%\noindent {\bf References}
%\input references.tex
\end{document}